\documentclass[12pt]{emulateapj} 
\usepackage{amsmath}
\usepackage{amssymb}
\usepackage{amstext}
\usepackage{graphicx}
\usepackage{epstopdf}
\usepackage{epsfig}
\usepackage{url}
\usepackage{color}
\definecolor{myColor}{rgb}{0.9,0.9,0.9}  

\submitted{Accepted for publication in The Astrophysical Journal Letters}

\begin{document}
\renewcommand\bottomfraction{.9}
\shorttitle{H$_2$O in hot Jupiter Atmospheres} 
\title{H$_2$O abundances in the atmospheres of three hot Jupiters}
\author{Nikku Madhusudhan\altaffilmark{1}, Nicolas Crouzet\altaffilmark{2,3}, Peter R. McCullough\altaffilmark{2,4}, Drake Deming \altaffilmark{5}, Christina Hedges\altaffilmark{1}}


\altaffiltext{1}{Institute of Astronomy, University of Cambridge, Cambridge CB3 0HA, United Kingdom {\tt nmadhu@ast.cam.ac.uk}}
\altaffiltext{2}{Space Telescope Science Institute, Baltimore, MD 21218, USA}
\altaffiltext{3}{Dunlap Institute for Astronomy \& Astrophysics, University of Toronto, 50 St. George Street, Toronto, Ontario, Canada M5S 3H4}
\altaffiltext{4}{Department of Physics and Astronomy, Johns Hopkins University, 3400 North Charles Street, Baltimore, MD 21218, USA}
\altaffiltext{5}{Department of Astronomy, University of Maryland, College Park, MD 20742, USA}

\begin{abstract}
The core accretion theory for giant planet formation predicts enrichment of elemental abundances in planetary envelopes caused by runaway accretion of planetesimals, which is consistent with measured super-solar abundances of  C, N, P, S, Xe, and Ar in Jupiter's atmosphere. However, the abundance of O which is expected to be the most dominant constituent of planetesimals is unknown for solar system giant planets, owing to the condensation of water in their ultra-cold atmospheres, thereby posing a key unknown in solar system formation. On the other hand, hundreds of extrasolar `hot Jupiters' are known with very high temperatures ($\gtrsim$ 1000 K) making them excellent targets to measure H$_2$O abundances and, hence, oxygen in their atmospheres. We constrain the atmospheric H$_2$O abundances in three hot Jupiters (HD~189733b, HD~209458b, and WASP-12b), spanning a wide temperature range (1200-2500 K), using their near-infrared transmission spectra obtained using the HST WFC3 instrument. We report conclusive measurements  of H$_2$O in HD~189733b and HD~209458b, while that in WASP-12b is not well constrained by present data. The data allow nearly solar as well as significantly sub-solar abundances in HD~189733b and WASP-12b. However, for HD~209458b, we report the most precise H$_2$O measurement in an exoplanet to date that suggests a $\sim20-135\times$ sub-solar H$_2$O abundance. We discuss the implications of our results on the formation conditions of hot Jupiters and on the likelihood of clouds in their atmospheres. Our results highlight the critical importance of high-precision spectra of hot Jupiters for deriving their H$_2$O abundances.
\end{abstract} 


\section{Introduction} 
\label{sec:intro}

Infrared observations in recent years are being used to place statistically significant constraints on the dominant molecular compositions of exoplanetary atmospheres. The most observed exoplanets to date are hot giant planets whose large scale-heights and high temperatures ($\sim$1200-3000 K) make them particularly conducive to atmospheric observations. Several prominent molecules containing carbon and oxygen and several atomic species are expected to be abundant and observable in these atmospheres, making giant exoplanets rich laboratories for understanding atmospheric chemistry (Madhusudhan et al 2014). 

Atmospheric elemental abundances of solar-system giant planets have led to important constraints on the origins of the solar system. The observed super-solar enrichments of C, S, N, and inert gases, support the formation of Jupiter by core-accretion (Owen 1999; Atreya \& Wong 2005). However, the oxygen abundance of Jupiter is yet unknown. The upper atmosphere of Jupiter (P $<$ 1 bar) has T $<$ 200 K, causing H$_2$O to condense and to be confined to the deepest layers ($>$ 10 bar), requiring dedicated probes to measure it. The upcoming Juno mission to Jupiter (Bolton 2010) aims to measure its O abundance, which is important to estimate the amount of water ice that was available in the planetesimals forming Jupiter and the rest of the solar system. The O/H and C/O ratios are easier to measure for hot giant exoplanets than they are for solar-system giant planets (Madhusudhan 2012). The vast majority of extrasolar gas giants known have equilibrium temperatures of $\sim$1000-3000 K, thus hosting gaseous H$_2$O in their atmospheres accessible to spectroscopic observations. 

Recently, major advancements in atmospheric spectroscopy of exoplanets are being made using the Wide Field Camera 3 (WFC3) on the Hubble Space Telescope ({\it HST}) (McCullough \& MacKenty 2012). The advantages of WFC3 in spatial scan mode allow extremely high precision in spectroscopic observations (e.g. Deming et al. 2013). Furthermore, the WFC3 bandpass ($\sim 1.1-1.7 \mu$m) contains a strong H$_2$O band at $\sim$1.4 $\mu$m providing a valuable tool for measuring H$_2$O abundances in exoplanetary atmospheres. Several recent studies have reported HST WFC3 spectra for a wide range of transiting exoplanets (e.g. Berta et al. 2012; Deming et al. 2013; Swain et al. 2013; Mandell et al. 2013; Sing et al. 2013; Kriedberg et al. 2014; Knutson et al. 2014; Ranjan et al. 2014). Such spectra are used to constrain the atmospheric compositions of the day-night terminator regions of transiting exoplanets using atmospheric models and retrieval methods (e.g. Madhusudhan \& Seager 2009). 

In the present work, we use HST WFC3 transmission spectra of three hot Jupiters to constrain the H$_2$O abundances in the day-night terminator regions of their atmospheres. In what follows, we explain our methodology in section~\ref{sec:method}. We present the constraints on the H$_2$O abundances of the planets in section~\ref{sec:results}. We discuss the implications of our results and future work in section~\ref{sec:discussion}. 

\section{Methodology}
\label{sec:method}

We examine WFC3 transmission spectra of three hot Jupiters: HD~189733b (McCullough et al. 2014), HD~209458b (Deming et al. 2013), and WASP-12b (Stevenson et al. 2014). We model the transmission spectra of these planets and retrieve the H$_2$O abundances from the spectral data using the exoplanet atmospheric retrieval method developed in Madhusudhan \& Seager (2009), Madhusudhan et al. (2011a), and Madhusudhan (2012). Since we are using transmission spectra, we model the atmosphere at the day-night terminator region of the planet (Madhusudhan \& Seager 2009). We consider a cloud-free plane-parallel atmosphere and compute line-by-line radiative transfer under the assumption of hydrostatic equilibrium and local thermodynamic equilibrium. The model atmosphere consists of 100 layers, in the pressure range of $10^{-6} - 100$ bar. We compute the net absorption of the stellar light  by the planetary atmosphere as the rays traverse the day-night terminator region of the spherical planet, appropriately integrating over the annulus. 

The atmospheric composition and temperature profile at the terminator are free parameters in the model. The parametric temperature profile is described in Madhusudhan \& Seager (2009), and has six free parameters. We include all the major opacity sources expected in hot Jupiter atmospheres, namely molecular opacity due to H$_2$O, CO, CH$_4$, CO$_2$, C$_2$H$_2$, HCN and H$_2$-H$_2$ collision-induced absorption (CIA) (Madhusudhan 2012; Moses et al. 2013), though H$_2$O contributes the most dominant opacity in the HST WFC3 G141 bandpass; the line-list data are described in Madhusudhan (2012). We assume uniform mixing ratios for all the   molecules in the atmosphere. We also allow for a free vertical offset in the data, accounting for uncertainties in the absolute level of the observed spectrum and in the base pressure level corresponding to the visible radius of the planet; we nominally assume a base pressure of 1 bar. Thus, overall, there are 13 free parameters in the model: six for the temperature profile, six molecules, and a free offset. 

We use a Bayesian approach to explore the model parameter space and derive posterior probability distributions of the model parameters given the data. For this purpose, we use the Markov chain Monte Carlo (MCMC) method with the Metropolis-Hasting scheme and a Gibbs sampler (see e.g. Madhusudhan et al. 2011a). We assume uniform priors for all the parameters, and explore a wide range of temperatures and chemical compositions for each planet. 
But, as discussed in section~\ref{sec:results} below, we obtain statistically significant constraints only on the H$_2$O abundance and we do not detect any other molecule, which is expected since H$_2$O is the dominant source of opacity in this bandpass. The temperature profiles span a wide range of temperatures well beyond those feasible in the observable atmospheres of the hot Jupiters considered; e.g. 400-2500 K for  HD~189733b and HD~209458b (Showman et al. 2009), and 1000-3000 K for WASP-12b (e.g. Spiegel et al. 2009). For reference, the equilibrium temperatures of  the planets, assuming efficient redistribution, are 1200 K (HD~189733b), 1450 K (HD~209458b) and 2500 K (WASP-12b).

\begin{figure}[t]
\centering
\includegraphics[width = 0.5\textwidth]{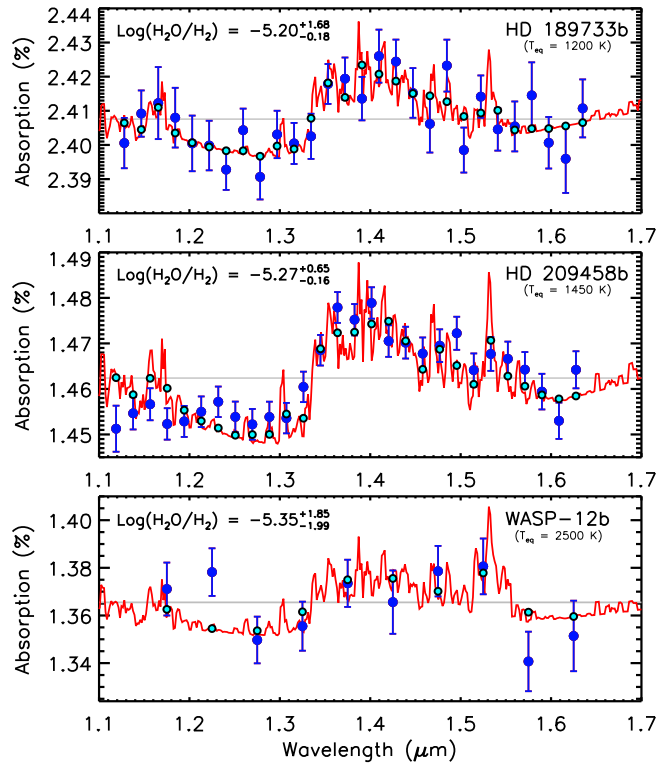}
\caption{Observations and model spectra of three hot-Jupiters. In each case, the blue circles with error bars show the data. The red curve is the best-fit model spectrum, and the cyan circles show the model binned to the same resolution as the data. The H$_2$O/H$_2$ volume mixing ratios in the models are: HD~189733b ($9\times10^{-6}$), HD~209458b ($2\times10^{-5}$), and WASP-12b ($3\times10^{-5}$). The gray horizontal line shows a best-fit featureless spectrum.} 
\label{fig:spectra}
\end{figure}

\section{Constraints on H$_2$O Abundances}
\label{sec:results}

We now present constraints on the atmospheric H$_2$O abundances for the three hot Jupiters in our sample: HD~189733b, HD~209458b, and WASP-12b. The motivation behind choosing these three planets is twofold. Firstly, the planets represent transiting hot Jupiters over a wide range of irradiation: HD~189733b (T$_{eq} \sim 1200$ K) is one of the coolest transiting hot Jupiters known and WASP~12b (T$_{eq} \sim 2500$ K) is amongst the hottest. Secondly, of all the hot Jupiters that have been observed using transmission spectroscopy with HST WFC3, these planets have the best spectroscopic precision. Consequently, the goal is to conduct a homogenous estimation of H$_2$O abundances in a diverse sample of hot Jupiters using the best observations with the same instrument. For each planet, we report constraints on the atmospheric H$_2$O volume mixing ratio, i.e. number fraction relative to H$_2$, at the day-night terminator, as well as the characteristic temperature (T$_{\rm phot}$) at the 100-mbar pressure level which typically corresponds to the planetary photosphere. We do not detect any other molecule in any of the planets in our sample. 

\subsection{HD~189733b}

We constrain the H$_2$O abundance at the terminator region of HD~189733b using the HST WFC3 transmission spectrum reported by McCullough et al. (2014). The observed spectrum and a best-fit model spectrum are shown in Fig.~\ref{fig:spectra}. The posterior probability distribution of the H$_2$O mixing ratio is shown in Fig.~\ref{fig:hist}. We constrain the H$_2$O mixing ratio to be $\log{\rm (H_2O/H_2)} = -5.20^{+1.68}_{-0.18}$, which corresponds to a 1-$\sigma$ range of $4.1 \times10^{-6}$ - $3.0 \times10^{-4}$, or 3-200$\times$ sub-solar for the corresponding temperature range; for $T \lesssim 1200$ K,  the H$_2$O mixing ratio for solar abundances is $\sim$$10^{-3}$, as shown in Fig.~\ref{fig:zplot} (also see Madhusudhan 2012). The central, most probable, value of $\sim 6.3 \times10^{-6}$ is $\sim$150$\times$ sub-solar. However, the data are also consistent with a solar abundance composition at  the 2-$\sigma$ level. The corresponding constraints on the representative temperature (T$_{\rm phot}$) is determined to be T$_{\rm phot}$ = $787^{+511}_{-47}$ K, which, as expected, is generally lower than the equilibrium temperature (T$_{eq}$) of the planet assuming efficient redistribution (T$_{eq} \sim 1200$ K).  Transmission spectra probe cooler regions of the atmosphere compared to dayside emission spectra  because (a) the former probe the day as well as the cooler night side at the terminator region, and (b) they probe higher altitudes (or lower pressures) and hence cooler regions in the atmosphere. 

Our estimate of the H$_2$O abundance in HD~189733b is the most stringent for this planet to date, and is lower compared to previous estimates using other transmission datasets. Swain et al. (2008) used a near-infrared transmission spectrum (1.4-2.4 $\mu$m) obtained with the {\it HST} NICMOS instrument to report a best-fitting model that had a H$_2$O mixing ratio of $5\times 10^{-4}$. Using the same dataset Madhusudhan \& Seager (2009) reported a similar mixing ratio, between 5 $\times 10^{-4}$ and 0.1.  However the observational uncertainties of the NICMOS spectra have been debated (Crouzet et al. 2012;  Gibson et al. 2012;  Waldmann et al. 2013; Swain et al., 2014). Constraints from broadband transit photometry have also been inconclusive (Tinetti et al. 2007; Desert et al. 2011). On the other hand, our estimate is consistent with upper-limits on the H$_2$O abundance in the dayside atmosphere of HD~189733b reported by various previous studies (e.g. Madhusudhan \& Seager 2009; Swain et al. 2009; Lee et al. 2012; Line et al. 2012). But again, the {\it Spitzer} photometry used to derive some of the previous abundances have since been revised (e.g. Knutson et al. 2012). 

\subsection{HD~209458b}

\begin{figure}[]
\centering
\includegraphics[width = 0.5\textwidth]{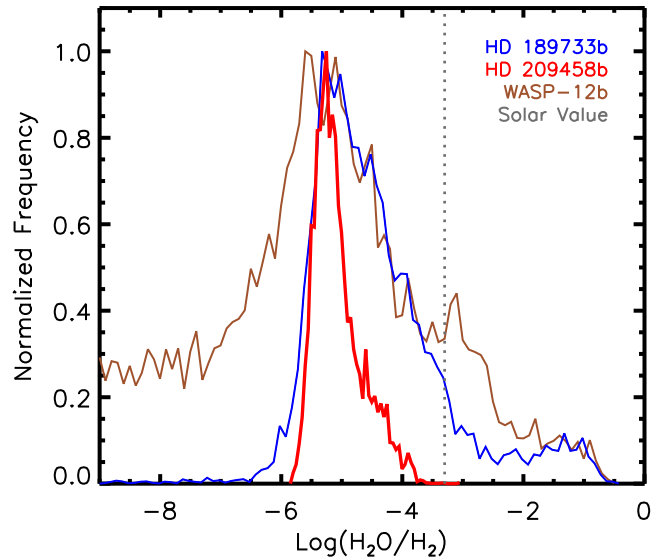}
\caption{Posterior probability distributions of H$_2$O abundance in three hot Jupiters (HD~189733b, HD~209458b, and WASP-12b) from model fits to the data.}
\label{fig:hist}
\end{figure}

For HD~209458b, we use the WFC3 transmission spectrum recently reported by Deming et al. (2013). In their study, using a model grid, Deming et al noted the apparently low H$_2$O abundance required to explain their data. In the present study, we quantify the H$_2$O abundance at the day-night terminator region of the planet using a more rigorous statistical retrieval. The observed and model spectra are shown in Fig.~\ref{fig:spectra}, and the posterior probability distribution of the H$_2$O mixing ratio is shown in Fig.~\ref{fig:hist}. We conclusively measure the H$_2$O abundance in the atmosphere, with a mixing ratio of $\log$(H$_2$O/H$_2$) = $-5.27^{+0.65}_{-0.16}$, i.e. a 1-sigma range of $3.7 \times10^{-6}$ - $2.4 \times10^{-5}$, which is $\sim$20-135 $\times$ sub-solar in the corresponding temperature range, as shown in Fig.~\ref{fig:zplot}. The central value of $5.4 \times 10^{-6}$, is $\sim$100$\times$ sub-solar. The data rule out a solar abundance H$_2$O at over 3-$\sigma$ confidence. The corresponding constraint on the representative temperature is T$_{\rm phot} = 1071^{+103}_{-234}$ K. 

Our estimate of the H$_2$O abundance of HD~209458b is the most stringent chemical constraint for an exoplanet to date, and is also consistent with previously reported upper limits. As mentioned above, our low estimate is consistent with the general conclusion of Deming et al. (2013) about a sub-solar H$_2$O abundance required to match the data. On the other hand, several previous studies have used multi-band broadband photometric observations which allowed upper-limits on the H$_2$O abundance of $\sim10^{-4}$ in the dayside atmosphere of the planet (e.g. Seager et al. 2005; Madhusudhan \& Seager 2009; Swain et al. 2009b). Previous studies using low-resolution transmission spectrophotometry of HD~209458b have suggested H$_2$O abundances of $\sim$10$^{-4}$-10$^{-3}$ (Barman  2007; Beaulieu et al. 2010) which are higher than our present measurement, but the statistical limits were not reported in those studies precluding a direct comparison with our results.  

\subsection{WASP-12b}
The observed spectrum of WASP-12b (Stevenson et al. 2014) and the best-fit model are shown in Fig.~\ref{fig:spectra}. The observational uncertainties in this case are larger compared to those of HD189733b and HD~209458b, because the latter two planets orbit much brighter host stars which lead to higher photon fluxes and hence better precisions; moreover,  WASP-12b was not observed in spatial scan mode of WFC3 unlike the other two systems. Consequently, our constraint on the H$_2$O abundance of WASP-12b is much less precise than those for HD189733b and HD~209458b. A wide range of H$_2$O abundances are permissible by the data. As shown in Fig.~\ref{fig:hist}, we estimate an H$_2$O mixing ratio of $\log{\rm (H_2O/H_2)} = -5.35^{+1.85}_{-1.99}$, corresponding to a 1-$\sigma$ range of ${\rm (H_2O/H_2)} = 4.5 \times10^{-8}$ - $3.1  \times10^{-4}$. The central value is $\sim$100 times sub-solar, though even at 1-$\sigma$ confidence the data permit a wide range of mixing ratios ranging from nearly solar to a non-detection, as shown in Fig.~\ref{fig:hist}; an H$_2$O mixing ratio below $\sim$10$^{-7}$ is indistinguishable  from zero with these data. The temperature constraint is T$_{\rm phot} = 1294^{+496}_{-50}$ K, which is only nominal since even a featureless spectrum fits the data relatively well. 

A potentially low H$_2$O abundance in the atmosphere of WASP-12b is consistent with previous studies. Several studies on the atmospheric composition at the terminator region of WASP-12b using transmission spectroscopy have found the various data sets to be fit equally well or equally poorly with oxygen-rich as well as carbon-rich models (e.g. Cowan et al. 2012; Swain et al. 2013; Mandell et al. 2013; Stevenson et al. 2014). On the other hand, observations of thermal emission obtained at occultation have been used to suggest a high C/O ratio (i.e. a carbon-rich composition) in the dayside atmosphere of the planet (Madhusudhan et al. 2011a; Madhusudhan 2012; but cf Crossfield et al. 2012). A high C/O ratio would naturally explain the low H$_2$O abundance we find, but higher precision WFC3 spectra will be required to support that hypothesis. In the future, joint constraints from both transmission and emission spectra will be able to better constrain the global H$_2$O abundance (e.g. Burrows et al. 2010; Griffith 2013). 

\section{Discussion} 
\label{sec:discussion}

\begin{figure}[]
\centering
\includegraphics[width = 0.5\textwidth]{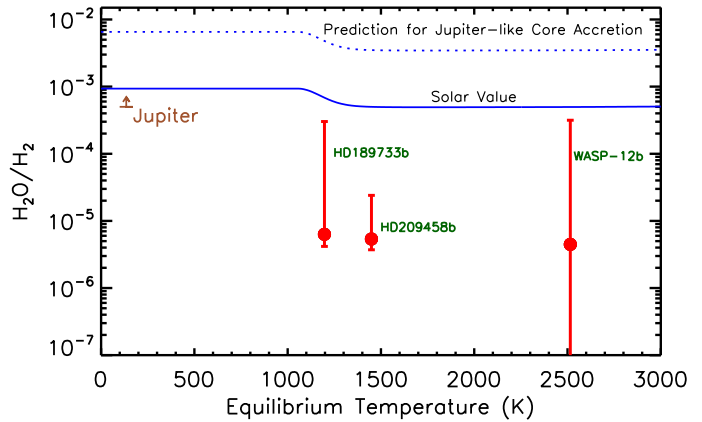}
\caption{Constraints on H$_2$O abundances. The red circles and error bars show the estimated mixing ratios with 1-$\sigma$ uncertainties derived from the posterior distributions shown in Fig.~\ref{fig:hist}.  The lower-limit on the H$_2$O mixing ratio in Jupiter is shown in brown (Atreya \& Wong 2005). The blue lines show H$_2$O mixing ratios expected assuming solar abundances (solid line) and formation of Jupiter by core accretion (dashed line); see section~\ref{sec:discussion}.}
\label{fig:zplot}
\end{figure}

Figure~\ref{fig:zplot} shows our H$_2$O estimates in comparison to solar abundance expectations. For a giant planet atmosphere of solar composition (Asplund et al. 2009), the H$_2$O/H$_2$ mixing ratio would  be $\sim$5$\times10^{-4}$ for  T$\gtrsim$1200 K, and up to $\sim$$10^{-3}$ for T$\lesssim$1200 K (Madhusudhan 2012). Our H$_2$O constraints for all the planets at 1-$\sigma$ suggest lower abundances than solar-abundance predictions, as shown in Fig.~\ref{fig:zplot}, though two of the planets (HD~189733b and WASP-12b) are consistent with solar composition at 2-$\sigma$. On the other hand, the H$_2$O abundance of HD~209458b is highly sub-solar, by 20-135 times. 

The low H$_2$O abundances we infer are surprising for two reasons. Firstly, the core-accretion theory for the formation of giant planets predicts enhancements in their atmospheric elemental abundances relative to their stellar abundances. The best example is Jupiter for which all the measured elements are 2-3$\times$ super-solar, though only a lower-limit is available for oxygen (Atreya \& Wong 2005). Assuming a solar composition nebula, current formation models predict $\sim$7$\times$ enhancement in Jupiter's oxygen abundance relative to solar (Mousis et al. 2012) leading to a corresponding enhancement in the H$_2$O abundance, as shown in Fig.~\ref{fig:zplot}. Secondly, H$_2$O is expected to be the most abundant oxygen-bearing molecule at all temperatures between 300-3500 K in a solar abundance atmosphere. In what follows, we discuss some possible explanations for low H$_2$O abundances in hot Jupiters. 

\subsection{Obscuration due to Clouds and Hazes}
\label{sec:clouds}

Several recent studies have suggested that the possible presence of clouds might be obscuring molecular features in transmission spectra of hot Jupiters. The earliest such suggestion was made by Charbonneau et al. (2002) to explain the lack of a strong Na feature in the optical transmission spectrum of HD~189733b. More recent observations in UV and visible indicate a featureless spectrum with a steep blue-ward slope, which has been suggested as evidence for high-altitude Rayleigh scattering  (Pont et al. 2013), and/or particulate scattering (Lecavelier des Etangs 2008). Alternatively, McCullough et al. (2014) interpret the same data with a clear planetary atmosphere and a spotted stellar photosphere. Using similar observations, Sing et al. (2013) also suggested the presence of hazes in the very hot Jupiter WASP-12b. Overall, the inferences of hazes/clouds from transmission spectra are motivated by the non-detection of spectral features of expected molecules and a blue-ward rise in the optical. On the other hand, observational constraints on the chemical composition of the haze/cloud material are non-existent. 

If clouds are indeed responsible for our low estimates of H$_2$O abundances, our results stress the need for rigorous theoretical efforts to explain several challenges in the cloud hypothesis. Firstly, within our reported constraints the H$_2$O abundance in HD~209458b (T$_{\rm eq}$$\sim$1450 K) can be lower compared to that in the cooler  HD~189733b ($\sim$ 1200 K), which is counterintuitive considering that the chances for clouds should increase with decreasing temperature; T$_{\rm eq}$ is the equilibrium temperature assuming fully efficient redistribution. 
Secondly, if clouds were responsible for the low inferred H$_2$O abundance in HD~209458b, considering even a solar abundance composition would require clouds at altitudes of 0.5 mbar (Deming et al. 2013). It is unclear if clouds can persist at such low pressures, i.e. high up in the atmosphere. Fortney et al (2010) suggest the possibility of clouds in hot Jupiters at pressures of several millibars, but not below a millibar. Furthermore, Spiegel et al. (2009) find that gravitational settling alone can preclude high-temperature condensates such as TiO from being abundant at low pressures ($\lesssim$10 mbar); refractory condensates (e.g. MgSiO$_3$) proposed for cloud composition in such atmospheres can be heavier than TiO making it even harder to keep them aloft. Alternately, it is also possible that the molecular features are obscured not by an opaque cloud deck but instead by a continuous opacity that is distributed uniformly with altitude, e.g. due to hazes (Deming et al. 2013; Pont et al. 2013). The composition and plausibility of hazes over the large temperature range of hot Jupiters needs to be investigated theoretically. The planets in our sample cover the entire range of hot Jupiter temperatures (T$_{eq} \sim$ 1200 - 2500 K), meaning that if the cloud hypothesis is correct it implies that clouds/hazes are ubiquitous in exoplanetary atmospheres.

\subsection{Possible Connection to Formation Conditions}
\label{sec:formation}

An alternative explanation to the low H$_2$O abundances we are inferring may be rooted in the formation conditions of the planet. A low H$_2$O abundance in a hot Jupiter atmosphere is possible in one of two ways: (a) either the overall metallicity is low, i.e. the oxygen abundance (O/H) is low, but the relative elemental ratios are solar (e.g. C/O = 0.5), or (b) the overall metallicity is solar or super-solar but the C/O ratio is high (e.g. C/O $\gtrsim$ 1) (Madhusudhan 2012). In either of the two cases, the implication is that substantially less H$_2$O ice accreted via planetesimals onto the forming planet. Similarly, the low Na and K abundances observed in hot Jupiter atmospheres which have thus far been attributed to clouds/hazes, as discussed above, may also be originating from their formation conditions. 

Relating atmospheric abundances to formation conditions of exoplanets are motivated by recent studies suggesting the possibility of measuring C/H, O/H, and C/O ratios in atmospheres of giant exoplanets (e.g. Madhusudhan et al. 2011a, Madhusudhan 2012). Several studies have suggested possible scenarios for H$_2$O depletion in giant planets. An early investigation into this question was pursued in the context of Jupiter in the solar-system for which a sub-solar H$_2$O abundance was reported by the Galileo probe (Owen et al. 1999; Atreya et al. 1999). A common explanation of the low H$_2$O abundance in Jupiter is that the probe landed in a dry spot (e.g. Atreya \& Wong 2005). However, Lodders (2004) suggested the possibility of Jupiter forming by accreting tar-dominated planetesimals instead of those dominant in water ice. 

More recently, following the suggestion of C/O $\geq$ 1 in WASP-12b (Madhusudhan et al. 2011a), several studies have suggested formation mechanisms that explain high C/O ratios in giant exoplanets. Oberg et al. (2011) suggested that C/O ratios in giant planetary envelopes depend on the formation location of the planets in the disk relative to the ice lines of major C and O bearing volatile species, such as H$_2$O, CO, and CO$_2$, since the C/O ratio of the gas approaches 1 outside the CO and CO$_2$ ice lines. By predominantly accreting such C-rich gas, more so than O-rich planetesimals, Gas giants could, in principle, host C-rich atmospheres even when orbiting O-rich stars. On the other hand, it may also be possible that inherent inhomogeneities in the C/O ratios of the disk itself may contribute to higher C/O ratios of the planets relative to the host stars (Kuchner \& Seager 2005; Madhusudhan et al. 2011b; Mousis et al. 2012; Moses et al. 2013; Ali-Dib et al. 2014).   

\acknowledgements{We acknowledge support in part from HST program GO 12881 provided by NASA through a grant from the Space Telescope Science Institute, which is operated by the Association of Universities for Research in Astronomy, Inc. We thank the anonymous referee for helpful comments. NM thanks David Spiegel, Richard Freedman, Jonathan Tennyson, Kevin Stevenson, Avi Mandell, Joseph Harrington, Jacob Bean, John Moriarty, and Debra Fischer for helpful discussions.} 
\newline

\end{document}